\documentclass[sigconf]{acmart}

\usepackage{subcaption}
\AtBeginDocument{%
  \providecommand\BibTeX{{%
    \normalfont B\kern-0.5em{\scshape i\kern-0.25em b}\kern-0.8em\TeX}}}

\copyrightyear{2024}
\acmYear{2024}
\setcopyright{rightsretained}
\acmConference[CHI EA '24]{Extended Abstracts of the CHI Conference on Human Factors in Computing Systems}{May 11--16, 2024}{Honolulu, HI, USA}
\acmBooktitle{Extended Abstracts of the CHI Conference on Human Factors in Computing Systems (CHI EA '24), May 11--16, 2024, Honolulu, HI, USA}
\acmDOI{10.1145/3613905.3650938}
\acmISBN{979-8-4007-0331-7/24/05}

\usepackage{todonotes}

\begin{document}


\title{3DA: Assessing 3D-Printed Electrodes for Measuring Electrodermal Activity}




\settopmatter{authorsperrow=4}

\author{Martin Schmitz}
\orcid{0000-0002-7332-3287}
\affiliation{
  \institution{Saarland University,\\ Saarland Informatics Campus}
  \city{Saarbr\"{u}cken}
  \country{Germany}
}
\email{mschmitz@cs.uni-saarland.de}

\author{Dominik Schön}
\orcid{0000-0003-2704-2852}
\affiliation{%
  \institution{Technical University of Darmstadt}
  \city{Darmstadt}
  \country{Germany}
}
\email{schoen@tk.tu-darmstadt.de}

\author{Henning Klagemann}
\orcid{0009-0001-7433-6506}
\affiliation{%
  \institution{Technical University of Darmstadt}
  \city{Darmstadt}
  \country{Germany}
}
\email{henning.klagemann@stud.tu-darmstadt.de}

\author{Thomas Kosch}
\orcid{0000-0001-6300-9035}
\affiliation{%
  \institution{HU Berlin}
  \city{Berlin}
  \country{Germany}}
\email{thomas.kosch@hu-berlin.de}

\renewcommand{\shortauthors}{Schmitz et al.}

\begin{abstract}
 Electrodermal activity (EDA) reflects changes in skin conductance, which are closely tied to human psychophysiological states. For example, EDA sensors can assess stress, cognitive workload, arousal, or other measures tied to the sympathetic nervous system for interactive human-centered applications. Yet, current limitations involve the complex attachment and proper skin contact with EDA sensors. This paper explores the concept of 3D printing electrodes for EDA measurements, integrating sensors into arbitrary 3D-printed objects, alleviating the need for complex assembly and attachment. We examine the adaptation of conventional EDA circuits for 3D-printed electrodes, assessing different electrode shapes and their impact on the sensing accuracy. A user study (N=6) revealed that 3D-printed electrodes can measure EDA with similar accuracy, suggesting larger contact areas for improved precision. We derive design implications to facilitate the integration of EDA sensors into 3D-printed devices to foster diverse integration into everyday objects for prototyping physiological interfaces.
\end{abstract}




\begin{teaserfigure}
\centering
  \includegraphics[width=\textwidth]{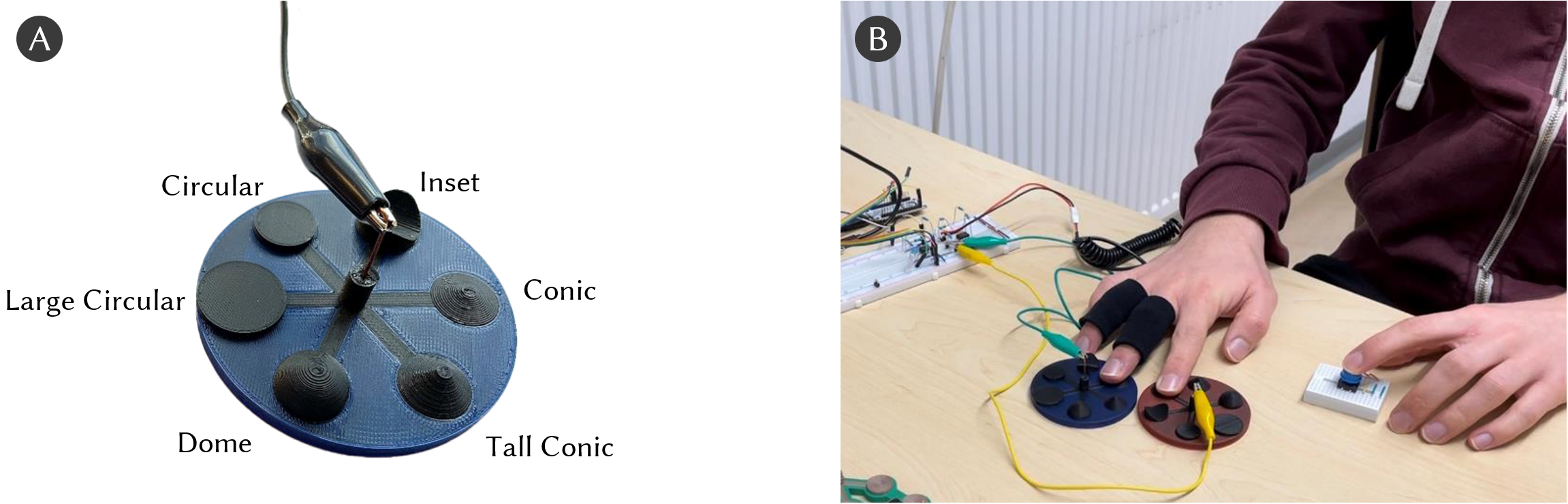}
  \caption{We evaluate the sensing efficiency of 3D-printed electrodes for measuring Electrodermal Activity (EDA). We propose six different electrode sizes and shapes (A) and compare them to a commercially available EDA sensor in a user study (B).
  }
  \label{fig:teaser}
\end{teaserfigure}


\maketitle

\section{Introduction}
Electrodermal Activity (EDA) sensors play an important role in measuring skin sweat, offering valuable insights into the psychophysiological states of users, such as stress~\cite{bhoja2020psychophysiological,Stuldreher_2020} or mental workload~\cite{10.1145/3582272}. The sympathetic nervous system~\cite{doi:10.1126/science.270.5236.644} dynamically adjusts sweat secretion in response to psychological conditions, allowing the inference of a user's mental state~\cite{VANDERMEE202152, 10.1145/3319499.3328230}. For instance, a sudden loud noise rapidly increases skin sweat, activating a ``fight-or-flight'' response~\cite{doi:10.1126/science.270.5236.644}. EDA sensors detect such changes by applying electrodes with small voltages to the skin (i.e., typically on the fingers~\cite{10.1145/3411764.3445370}) and assessing its resistance, which decreases with higher sweat conductivity. Sensing EDA has interested the HCI community to implicitly assess user states, such as mental workload~\cite{10.1145/3582272}, stress~\cite{LIU201850}, or emotions~\cite{8653316}. Such measures can be used as implicit metrics for assessing user experience~\cite{10.1145/3319499.3328230} or providing adaptive user interfaces~\cite{10.1145/3604243}. 
Yet, the practical integration of EDA into user interfaces remains a challenge. While commercial devices are affordable, they require attaching electrodes to the user (i.e., to the fingers or feet), hindering the comfort and user acceptance of using EDA in interfaces. While previous work showed that electrodes could be integrated into common skin contact points, such as VR controllers~\cite{10.1145/3604270}, such integrations are lavishly and only cover specific use cases, demanding a more flexible approach to sense EDA in everyday interaction scenarios. In addition, depending on where such electrodes are integrated into a 3D object, they will also have different shapes, possibly changing their sensing characteristics.

This paper, therefore, explores the feasibility of 3D printing EDA sensors (3DA) to overcome these challenges and embed them seamlessly into everyday objects. To this end, we evaluate the suitability of differently shaped 3D-printed conductive electrodes for measuring EDA (see \autoref{fig:teaser}). We contribute a user study that compares the measured values for different electrode shapes with commercially available EDA sensors for six participants performing an auditory oddball task. 
Our results show that specific shapes achieve a solid correlation with the commercial EDA sensor, while other shapes show a low correlation. Based on the results, we discuss the effects of electrode shapes and materials on accuracy and conclude with design suggestions for integrating physiological EDA sensing into 3D-printed objects.

\section{Related Work}
Our work is rooted in sensing and using EDA for interactive applications. We, therefore, summarize related works in 3D-printed physiological sensors and using EDA for interactive applications.

\subsection{Sensing Electrodermal Activity}
The terminology and framework presented in Boucsein's work~\cite{boucsein2012electrodermal} constitute the foundation of EDA measurements. Boucsein's work summarizes how EDA works and offers detailed insights into various EDA applications. The methodology aligns with the recommendations proposed by Fowles et al.~\cite{fowles1981committee} and the subsequent 2012 report from the Society for Psychophysiological Research Ad Hoc Committee on Electrodermal Measures~\cite{https://doi.org/10.1111/j.1469-8986.2012.01384.x}. These guidelines inform our approach to measuring EDA and interpreting the resulting data in this research. In this context, recent EDA research has been dedicated to developing recording devices suitable for unobtrusive and ambulatory use outside traditional laboratory and medical settings~\cite{conchell2018design, s19112450, affanni2014wearable, SCHMIDT201696, 5415607}. These studies focus on creating EDA sensors with minimal hardware requirements, often utilizing small sensor chips. An emphasis is placed on affordability, enabling long-term measurements, and maintaining a certain measurement quality. Many works focus on wearable devices, such as wristbands and smartwatches, with considerations for alternative recording sites to enhance user comfort and acceptability~\cite{https://doi.org/10.1111/psyp.13645}.

\subsection{3D-Printed Physiological Sensors}
Extensive research has been conducted and continues in 3D-printed sensor structures.
Xu et al.~\cite{s17051166} conducted a comprehensive review, encompassing recent examples of sensors measuring physical variables such as temperature and pressure, interactive tactile and strain sensors, and physiological sensors, including EEG sensors. Furthermore, Dijkschoorn et al. have provided a detailed overview of techniques employed in the 3D printing sensor structures~\cite{jsss-7-169-2018}. For example, 3D-printed sensors can provide touch detection on customized objects~\cite{10.1145/3411764.3445595, 10.1145/3411763.3451536,Schmitz2019,Schmitz2021}.

3D-printed EDA sensors were rarely the objective of past research. Zhao et al.~\cite{10.1145/3544794.3558479} demonstrated the feasibility of creating a fully printed EDA sensor chip using graphene ink for wrist attachment. However, this comes with challenges such as material specificity and shape limitations, utilizing a rectangular electrode shape of 16 mm$^2$. Ho et al. focused on 3D printing electrodes, utilizing high-resolution powder bed printing for flexible, conductive structures in wearable sensors~\cite{https://doi.org/10.1002/advs.201902521}. They created interconnected porous structures using sugar grains as powder, requiring coating and optional conductive material filling. While their approach showed promising results for various physiological sensors, including EDA, it involves intricate steps. Further investigations are necessary, especially using FDM printing for easy integration into arbitrary objects. 

Numerous studies investigate 3D-printed electrodes, yet they typically fall into two categories: those that generally analyze 3D-printed electrodes for electrochemical sensing, utilizing methods such as Fused Deposition Modeling (FDM) and electrodeposition~\cite{ROHAIZAD2019104,KATSELI2019100}, and those that concentrate on specific sensors, employing techniques such as FDM~\cite{s20154292} or PolyJet printing~\cite{SALVO201296}. Alsharif et al. recently reviewed 3D-printed electrodes, specifically those not reliant on electrolytic gels~\cite{https://doi.org/10.1002/admt.202201677}. When applying electrodes for EDA, insights from conventional EDA electrodes, primarily silver chloride electrodes, can be employed. Studies on these electrodes highlight the correlation between the size of the contact area and measured skin conductance. Mahon and Iacono~\cite{https://doi.org/10.1111/j.1469-8986.1987.tb00280.x} observed a linear increase in the Skin Conductance Level (SCL) and Skin Conductance Response (SCR) amplitudes with a larger contact area. Thus, the electrode shape influences the sensing quality.

\subsection{Using Electrodermal Activity for Interactive Applications}
Previous HCI studies investigated EDA for interactive applications. For instance, Pan et al.~\cite{10.1145/1979742.1979716} proposed utilizing orienting responses as an ``implicit communication paradigm,'' employing SCRs to indicate interruptions during audio stream consumption. Kosch et al.~\cite{10.1145/3319499.3328230} demonstrated the utility of EDA as a measure for adaptive augmented reality production assistance systems. Shi et al.~\cite{10.1145/1240866.1241057} examined EDA as a cognitive load indicator in traffic control management tasks, correlating higher SCLs with increased task complexity. Klarkowski et al.~\cite{10.1145/3242671.3242714} explored the relationship between challenges in video games and EDA to enhance player experience. Furthermore, EDA is frequently passively recorded for subsequent analysis. Ayzenberg et al.~\cite{10.1145/2212776.2223802} developed a system to implicitly log users' stress levels during daily social interactions on mobile phones, aiming to aid users in recalling and managing stressful situations. Additionally, EDA is employed to capture emotional states, as demonstrated by Kim et al.~\cite{10.1145/3460418.3479356}, who designed glasses equipped with an EDA sensor to track emotions during story reading. However, with the emerging number of applications for interactive EDA, it becomes increasingly important to ensure scientifically correct EDA measures. Babaei et al.~\cite{10.1145/3411764.3445370} summarized previous research for interactive EDA, stated critique regarding their sensing approach, and provided directions on how the HCI community should assess EDA measures in the future.

Previous research showed that EDA can be used for interactive applications. Yet, integrating EDA electrodes into everyday objects remains a research challenge. Inspired by earlier work looking at EDA electrode shapes~\cite{https://doi.org/10.1111/j.1469-8986.1987.tb00280.x}, this work, therefore, assesses the measurement performance of customizable 3D-printed electrodes.

\section{Efficiency of 3D-Printed Electrode Shape on EDA Measurements}
We compare the EDA sensing performance of 3D-printed and commercially available EDA electrodes. We manipulate the electrode shape of the 3D-printed electrodes.

\subsection{Study Setup}
To evaluate the impact of electrode shape on measurement outcomes, we utilized 3D printing to create a diverse set of six electrode pairs (see~\autoref{fig:teaser}). We printed \emph{flat} circular electrodes since most commercially available electrodes are typically circular. Additionally, we produced electrodes with a larger diameter (\emph{large circular}) to investigate the influence of electrode area. To explore the use of non-flat, sloped electrodes, we printed four additional pairs, including a pair with a spherical shape (\emph{dome}). Two electrodes were \emph{conical} with flattened tips and varying heights, creating one pair with a more pointed configuration than the other. The final sloped electrodes were not curved outward but inward (\emph{inset}), providing the added benefit of guiding fingers intuitively to the correct location. 
The electrodes were printed using Protopasta Conductive PLA\footnote{\url{https://proto-pasta.com/pages/conductive-pla} -- last accessed 2024-03-20}, a composite material of PLA and carbon black, with a conductivity of up to 3.3 S/m (approximately 30 $\Omega$/cm). 

As a reference, we use the commercially available nickel dry Grove EDA electrodes by Seeed Technology\footnote{\url{https://wiki.seeedstudio.com/Grove-GSR_Sensor} -- last accessed 2024-03-20}, which are circular and 15 mm in diameter. Therefore, all 3D-printed electrodes, except the large circular ones, were printed with the same diameter. The electrodes were 3D-printed using an Original Prusa i3 MK3S printer with Multi Material Upgrade 2S (MMU2S), and the layer height was set to 0.2 mm with 100\% infill density using the rectilinear fill pattern from PrusaSlicer. All pipes printed of conductive filament had a cross-section of 4 mm\,$\times$\,4 mm and a length of 32.5 mm, extending from the outer end of the electrode to the center point of the board, where a lead was inserted into a cylinder printed on top of the pipes by heating the lead with a soldering iron.

To evaluate the performance of 3D-printed EDA sensors, it is also necessary to adapt existing EDA sensor circuits for compatibility with 3D-printed electrodes. More advanced signal processing tasks such as amplification and filtering can be optionally added as supplementary hardware components or executed through software. We intentionally focus on exosomatic recording using Direct Current (DC)~\cite{boucsein2012electrodermal} since it is most suitable for seamlessly integrating an EDA sensor into small objects. This approach eliminates the need for a reference electrode at a remote location and avoids the handling of Alternate Current (AC), for example, by using a DC-to-AC converter. 

\subsection{Method}
We utilized an oddball task, where brief tones are repeatedly presented audibly to the participant, with certain tones having different frequencies, referred to as oddballs~\cite{SQUIRES1975387} and initially utilized in neuroscience research to elicit a P300, an event-related potential detected by electroencephalography in response to perceiving an oddball stimulus~\cite{picton1992p300}. This task has been observed to also induce SCRs concurrently with the P300 known from neuroscience~\cite{https://doi.org/10.1111/j.1469-8986.1976.tb03076.x,  https://doi.org/10.1111/j.1469-8986.2010.01057.x}. Thus, the oddball task is well-suited for eliciting SCRs. It is important to note that they should be treated separately from assessing implicit EDA, denoting variations in skin conductance that are not intentionally provoked by a stimulus. The study employs a within-subjects design.

\subsubsection{Independent Variables} As a single independent variable, We investigate six electrode shapes: \textit{conic, tall conic, circular, large circular, dome, inset} (see~\autoref{fig:teaser}). 
We compare these to the Seeed Grove EDA electrode measurements during the oddball task.

\subsubsection{Dependent Variables}
We measure the skin conductance from the 3D-printed and commercially available Grove EDA electrodes using Siemens, a standard measure for electrical conductance. The conductance of the 3D-printed and Grove electrodes was measured simultaneously, and the sampling rate was synchronized between both sensors.

\subsection{Procedure \& Task}
The study involved six individuals (three identified as females and three as males). These volunteers were between 20 and 24 years old, with a mean age of 22 (SD = 1.3). Two participants reported consuming caffeine within the six hours preceding their involvement. The environmental conditions during the study were maintained at an average temperature of 21.4 $^{\circ}$C (SD = 1.3) and an average humidity of 45.4\% (SD = 2.8\%). We complied with the recommendations by \citet{10.1145/3411764.3445370} for conducting EDA studies. Upon arrival, participants were instructed to wash their hands with cold water. Subsequently, they sat at a table where the sensor and test electrodes were positioned. Detailed information about the study procedure and data collection was provided to them. Following the explanation, participants signed their informed consent. They were directed to attach the Grove reference electrodes to the middle phalanges of the ring and middle finger of the participant's non-dominant hand. 

The electrode connected to the positive potential of the twin EDA sensor was affixed to the middle phalanx of the middle finger. In contrast, the other electrode was attached to the middle phalanx of the ring finger. No electrolyte gel was used, as its application cannot be assumed for the specific type of EDA sensors under investigation in ambulatory settings. Participants were in contact with the electrodes for at least five minutes before the experiment started. Subsequently, the EDA recording was conducted iteratively with each pair of the six test electrodes. The procedure involved the participant placing the distal phalanx of their middle finger on the first printed test electrode connected to the positive potential of the twin EDA sensor. Simultaneously, the distal phalanx of the index finger was placed on the second corresponding printed electrode on the other board. A relaxation task was conducted as a baseline, during which participants were instructed to relax for a few minutes while the EDA recording commenced. The task was completed, and the recording stopped when the EDA curves of the two electrode pairs had approached a mutually consistent linear trend or a stable value, according to visual assessment.
Then, the participants engaged in a breathing task, where they were told to take a fast, deep breath, hold it for a second, and breathe out calmly. This procedure was repeated a second time, and the recording was stopped four to six seconds later. Then, the participant was asked to take a finger-sized button provided on the table with their dominant hand, which was not attached to any electrodes. The participant was instructed that an audio track would be played on which tones of two different pitches would be heard. The track would start with deep tones; whenever a high tone appeared, the participant was asked to press the button as fast as possible. In later iterations, the participant was reminded to press the button as soon as possible after the tone to increase their attention. The track was the same for all iterations and contained three tones of the higher pitch occurring after 13.5, 21.0, and 34.5 seconds. The low tone was a 1000 Hz tone, while the oddballs were 1500 Hz tones, all with a duration of 80 ms occurring at an interval of 1500 ms. EDA was recorded during all conditions. The electrodes were switched after each condition. The task was completed for all participants, and the entire procedure took approximately 60 minutes per individual.

\subsection{Data Processing}
\begin{figure}
    \centering
    \includegraphics[width=\columnwidth]{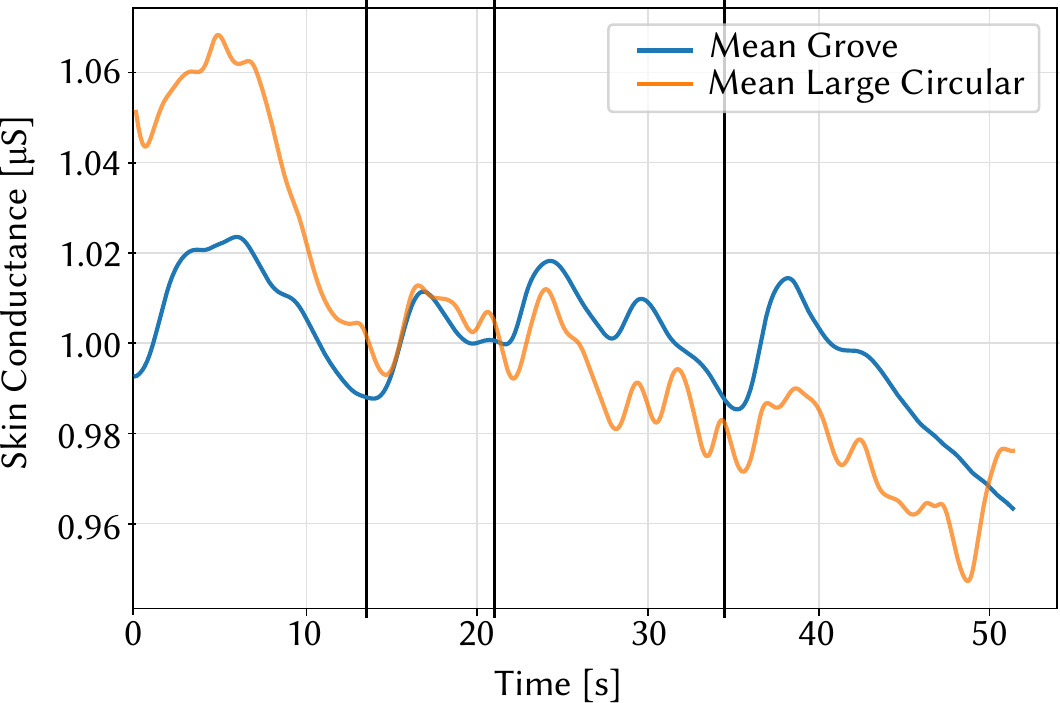}
    \caption{Exemplarily plot between the Grove EDA electrodes and the large circular 3D-printed electrodes with the highest correlation, averaged across all participants. The curves were divided by their mean and filtered with a low-pass filter at a cutoff frequency of 0.5 Hz. The vertical black lines indicate the times of the oddball stimulus.}
    \label{fig:sample_plot_mean_circular}
\end{figure}

We plotted the curves of all participants for all electrode pairs and all tasks after applying a Butterworth lowpass filter with a 0.5 Hz cutoff frequency. We chose this frequency to filter out as much noise as possible while still maintaining SCRs, which have a mean ascent time slightly more significant than two seconds~\cite{boucsein2012electrodermal}. Hence, a cutoff frequency of 0.5 Hz should preserve these changes. Besides, we divided the curves by their mean before plotting to map them all in a comparable range while keeping their variance. For the oddball task, we marked the times of occurring oddballs with vertical lines (see~\autoref{fig:sample_plot_mean_circular}).

We used cvxEDA to split the EDA curves into phasic and tonic components for visual inspection only. cvxEDA takes a parameter called alpha, a ``penalization for the sparse SMNA (sudomotor nerve activity) driver''\footnote{\url{https://github.com/lciti/cvxEDA} -- last accessed 2024-03-20}, which controls how many deflections in the curve are classified as SCRs. We set this parameter to 0.16 as this value seemed to achieve the best results for the curves in our plots. All other parameters were set to their default value. Since cvxEDA does not require any data preparation before its execution, we did not apply a lowpass filter for this step. However, following the developer's suggestion, we performed standard score normalization beforehand. We included the raw, normalized curves in translucent colors. We also used the NeuroKit2 library\footnote{\url{https://github.com/neuropsychology/NeuroKit} -- last accessed 2024-03-20} to separate phasic and tonic components from the EDA data measured during the oddball task. NeuroKit2 also detects SCR onsets and peaks, which we will use to analyze the oddball task further. We again used a lowpass filter with a 0.5 Hz cutoff frequency to plot this data. We divided the curves by their mean before passing the measurements to NeuroKit2 to obtain comparable results. The determined phasic components were additionally standard score normalized before plotting for better comparability. We calculated the Pearson correlation between the commercial and 3D-printed electrodes.

\subsection{Results}

\subsubsection{General Observations}
We observed that the initial skin conductance values measured by all printed test electrodes were high and exhibited a decreasing trend in the first 20-30 seconds, following a decaying curve. 
Beyond 30-40 seconds, the Protopasta test electrode curves appeared to align with the corresponding reference curves, as evidenced by simultaneous peaks in both.  In the context of the breathing task, we frequently identified two prominent peaks in the reference curves, consistent with expectations after each breath. Additional higher-frequency fluctuations accompanied these peaks. In many cases, the Protopasta curves mirror these major peaks.  The oddball task curves displayed increased fluctuations, with peaks often appearing seconds after the marked oddball times, aligning with our anticipated orienting responses. These peaks were also evident in the first differences plots, although with more fluctuations (see~\autoref{fig:sample_plot_mean_circular}).

\begin{table}[b]
\caption{Pearson correlations between the different electrode shapes and the commercial EDA sensor. Bold values indicate the highest correlations.}
\resizebox{\linewidth}{!}{
\begin{tabular}{lllllll}
\toprule
\textbf{Electrode}  & Circular & Large Circular & Dome & Conic & Tall Conic & Inset \\
\midrule
\textbf{Pearson's r} & .651     & \textbf{.683}           & .279 & .331 & \textbf{.680}          & .597  \\    
\bottomrule
\end{tabular}
}
\label{tab:results_corr}
\end{table}

\subsubsection{Correlation Between 3D-Printed and Commercial Electrodes}

\autoref{tab:results_corr} summarizes the results of the Pearson correlation between the six different 3D-printed electrode shapes and the Grove EDA electrodes (see~\autoref{fig:teaser}). We use the Pearson correlation to obtain a similarity metric between the SCL of both signals. Large circular electrodes exhibited the highest correlation values for these metrics. Pearson’s r, utilized to measure the resemblance of the test electrode curves to the reference curves, reached its peak with large circular electrodes (0.397 for breathing and 0.683 for the oddball task). The conic and dome electrodes achieved high r-values in the breathing task, while circular and tall conic electrodes secured high r-values in the oddball task. All 3D-printed electrodes yielded positive r-values in both tasks.


\subsubsection{EDA Classification}
We quantify how many of the detected SCRs were correct (i.e., precision), how many SCRs that occurred were detected (i.e., recall),  and calculate the F1 using precision and recall. \autoref{tab:results_classification} summarizes the findings. 
\begin{table}[b]
\caption{We measure the accuracy of identified SCRs by assessing precision, which indicates the proportion of correct detections. We also evaluate recall, which signifies the proportion of actual SCRs that were successfully detected. The F1 score, derived from precision and recall, is then calculated. Bold values indicate the highest classification values.}
\resizebox{\linewidth}{!}{
\begin{tabular}{lllllll}
\toprule
Electrode & Circular & Large Circular & Dome & Conic & Tall Conic & Inset \\
\midrule
Precision & .430     & \textbf{.437}           & .393 & .414 & .378             & .385  \\
Recall    & .475     & .735           & .561 & .594 & \textbf{.903}             & .696  \\
F1 Score  & .415     & \textbf{.608}           & .420 & .483 & .505              & .481 \\
\bottomrule
\end{tabular}
}
\label{tab:results_classification}
\end{table}
Following the computation of the mean across a broad spectrum of frequencies, all electrodes demonstrated similar precision. However, notable distinctions emerged in recall values: tall conic electrodes exhibited significantly high recall (0.903). Subsequently, large circular and inset electrodes recorded the following highest recall values. Large circular electrodes also attained the highest F1 score (0.608). Circular and dome electrodes registered the lowest F1 score. Observations regarding latency revealed that the smaller conic and dome electrode SCRs exhibited the most prolonged latency, while circular seemed to capture SCRs earlier than the reference electrodes. Regarding SCR amplitudes, tall conic electrodes measured the largest, while circular electrodes measured the lowest. After normalizing curves by their means, all test electrodes measured larger SCRs than the reference electrodes on average over all measured SCRs.

\section{Discussion}
We have proposed a method to incorporate EDA sensors into 3D-printed objects. Our focus in this research has been on employing simple measurement techniques, specifically the quasi-constant current and voltage methods, as well as the constant current and voltage method using DC. Two primary reasons guided our choice of these methods. Firstly, they are likely to be seamlessly integrated into devices with limited space requirements, such as most mobile devices. Secondly, these methods provide a foundation for further development. Our approach involved using circuits to measure the output voltage with a microcontroller. Alternatively, analog signal processing could be implemented in hardware, as Zhao et al.~\cite{10.1145/3544794.3558479} demonstrated, who employed binary signal classification. Similar to our study, more specialized data preparation circuitry could be added, where we incorporated amplification, such as segregating into phasic and tonic components. Examining sensor designs and conducting studies should be considered a proof of concept, offering a starting point for future investigations into integrating EDA sensors into 3D prints or for practical hands-on projects. 

We propose the following recommendations regarding selecting electrode shapes based on our results: 
It is advisable to utilize electrodes with a more extensive skin contact area to achieve improved accuracy for SCRs. 
A flat-shaped electrode is recommended when low contact resistances are a priority. 
A pointed shape may be employed if the objective is to attain heightened sensitivity. 
In cases where robustness is critical, particularly in applications where EDA recording may be susceptible to motion interference, electrodes with an inwardly curved shape prove advantageous because they yield stable results with reasonably high accuracy and provide a secure grip due to their geometry. Additionally, it is essential to consider sufficient time for polarization~\cite{10.1145/3411764.3445370}. Despite the challenges we encountered during our research, we envision that future conductive materials will improve their conductivity. At the same time, existing 3D printing fabrication pipelines~\cite{10.1145/3526114.3558719} will consider the option of tracking EDA in their algorithms.

\section{Conclusion and Future Work}
This paper investigates the application of 3D-printed electrodes for measuring Electrodermal Activity (EDA) and assesses how electrode shape impacts measurement accuracy. In a user study with six participants, we tested six 3D-printed electrodes with a conductive filament compared to commercially available off-the-shelf nickel electrodes. The results show that printed electrodes maintain EDA measurement accuracy, with the best performance from electrodes with a flat shape and a large diameter.  The findings suggest potential applications in everyday objects such as tools, devices, cups, steering wheels, and wristbands, facilitating ambulatory EDA recording without cables or adhesive electrodes. The paper's contribution paves the way for integrating EDA sensors into various everyday objects through 3D printing, enabling broader accessibility and application of EDA for interaction~\cite{10.1145/3604270} or biofeedback~\cite{sehrt2024closing}. Our work suggests future research directions with larger participant pools, including exploring other materials for printed electrodes to improve the accuracy, optimizing algorithm parameters for specific electrode shapes, testing its robustness against artifacts (e.g., motion), the influence of long-term use, and designing systems utilizing EDA as an input modality for arousal or stress.

\begin{acks}
This work has been funded by the German Federal Ministry of Education and Research (01IS17050), the European Union (European Regional Development Fund, ERDF), and the state of Saarland, Germany (project "Multi-Immerse").
\end{acks}
\newpage
\bibliographystyle{ACM-Reference-Format}
\bibliography{main}


\end{document}